\documentclass{appolb}
\usepackage{epsfig}
\def\PLB{{ Phys. Lett.}  B}
\def\NPA{{ Nucl. Phys.}  A}

\def\PRL{ Phys. Rev. Lett.}
\def\PRC{{ Phys. Rev.} C}

\def\Journal#1#2#3#4{{#1} {\bf #2}, #3 (#4)}

\begin{document}
\title{Azimuthally-sensitive interferometry and the source lifetime at RHIC
\thanks{Presented at 
the XXXIII International Symposium on Multiparticle Dynamics (ISMD2003),
Krak\'{o}w, Poland, 5-11 Sept 2003.}%
}
\author{Michael Annan Lisa, for the STAR Collaboration
\address{The Ohio State University, Physics Department, Columbus, OH 43210, USA}
%\and
%The STAR Collaboration
%\address{http://www.star.bnl.gov}
}
\maketitle
\begin{abstract}
Pion interferometry (``HBT'') measurements relative to the reaction plane
provide an estimate of the transverse source anisotropy at freeze-out,
which probes the system dynamics and evolution duration.  Measurements
by the STAR Collaboration indicate that the source is extended increasingly
out-of-plane with increasing impact parameter, suggesting a short evolution
duration
% of $\sim9$~fm/c,
 roughly consistent with estimates based on
azimuthally-integrated HBT measurements.
\end{abstract}
  
\section{Introduction}

Two-particle intensity interferometry (Hanbury-Brown--Twiss, or HBT) analyses of relativistic heavy ion
collisions represent the most direct experimental probes of the space-time structure of the source
generated in such collisions~\cite{HBTgeneral}.
At the Relativistic Heavy 
Ion Collider (RHIC), identical-pion HBT studies in Au+Au collisions~\cite{STARy1,PHENIXy1,MercedesQM02} 
yielded an apparent source size quantitatively consistent with measurements at lower energies, in contrast 
to predictions of larger sources based on QGP formation~\cite{RischkeGyulassy}.  In addition, hydrodynamical models, 
successful at RHIC in describing momentum-space quantities such as transverse momentum spectra and elliptic 
flow~\cite{starV2Prl}, have failed to reproduce the small HBT radii~\cite{heinzKolbWW02}.  This 
``HBT puzzle"~\cite{heinzPanic02,dumitru02} may arise because the system's lifetime is
shorter than predicted by models.

In non-central collisions, the initial anisotropic collision geometry generates greater transverse pressure gradients in
the reaction plane than perpendicular to it.  This leads to
preferential in-plane expansion (elliptic flow)~\cite{KSH00,voloshinQM,starV2Prl,olli} which diminishes the initial
spatial anisotropy.  The final (freeze-out) source shape should be sensitive to the evolution of the pressure
gradients and the system lifetime; a long-lived source would be less out-of-plane extended and perhaps in-plane extended.
Hydrodynamic calculations ~\cite{kolbQM02} predict a strong sensitivity of the HBT
parameters to the early source conditions and show that, while 
the source may still be {\it out-of-plane} extended after 
hydrodynamic evolution, a subsequent rescattering phase \cite{teaney} tends to make the source {\it in-plane}.  Thus,
the freeze-out source shape might discriminate between scenarios of the system's evolution.

Here, we present results of a measurement of azimuthally-sensitive HBT in Au+Au collisions at $\sqrt{s_{NN}}=200$~GeV.  We
discuss how these results may be used to extract, in a model-dependent way, the geometric anisotropy of the source at freeze-out.
Finally, we use this extracted geometry to estimate the evolution timescale of the system.

\section{Measurement}

The measurements were made using the STAR detector~\cite{starNim} at RHIC.  Particle trajectories and momenta were reconstructed using a Time Projection 
Chamber (TPC) with full azimuthal coverage, located inside a 0.5 Tesla solenoidal magnet. 
Au+Au event samples of 0--5\%, 5--10\%, 10--20\%, 20--30\% and 30--80\% of the total
cross hadronic section (based on charged particle multiplicity) are presented here.
The second-order event plane angle $\Psi_2$~\cite{poskanzer} for each
event was determined from the weighted sum of primary charged-particle
transverse momenta ($\vec{p}_T$)~\cite{STARfirstv2}.
Within the resolution which we determine from the random subevent
method~\cite{poskanzer}, $\Psi_2 \approx \Psi_{rp}$ (true reaction plane angle)
or~$\Psi_2 \approx \Psi_{rp}+\pi$;
{\it i.e.} the direction of the impact parameter vector is determined up to
a sign~\cite{poskanzer}, and the measured event plane is roughly
coplanar with the true reaction plane~\cite{STARv1v4}.

Pions were
selected according to their specific energy loss ($dE/dx$) in the TPC in the rapidity range $|y|<0.5$.  
Pair-wise cuts to remove two-track merging and single-track splitting effects were applied to signal and
background distributions~\cite{STARy1}.

Pairs of like-sign pions were placed into bins of 
$\Phi = \phi_{\rm{pair}} - \Psi_{2}$, where $\phi_{\rm{pair}}$ is the
azimuthal angle of the total pair momentum (${\mathbf k}={\mathbf p_1}+{\mathbf p_2}$).
Because we use the 2$^{\rm nd}$-order event plane, $\Phi$ is only defined in the range $(0,\pi)$.
For each bin, a three-dimensional 
correlation function is constructed in the ``out-side-long" decomposition~\cite{pb} 
of the relative momentum ${\mathbf q}$. 
% The numerator of the correlation function contains 
%pairs of pions from the same event, and the denominator contains pairs of pions from different events
%which have similar primary vertex position, reaction plane orientation, multiplicity, and magnetic field 
%orientation.
%$\pi^-$ pairs and $\pi^+$ pairs were mixed separately due to 
%charge-dependent acceptances but are combined to increase statistics; separate $\pi^+$ and $\pi^-$ analyses showed no significant differences.

Finite reaction plane resolution and finite width of the $\Phi$ bins has the effect of reducing the measured oscillation amplitudes.
A model-independent correction procedure \cite{HHLW02} is applied to each ${\mathbf q}$-bin in the numerator and 
denominator of each correlation function, the overall effect of which is to increase the 
amplitude of the oscillations of the HBT radii vs.~$\Phi$  ($\sim10-30\%$).

To account for the effect of final-state Coulomb interactions, we adopt the approach first proposed by Bowler~\cite{bowler} and
Sinyukov~\cite{sinyukov} and recently advocated by the CERES collaboration~\cite{ceresNpa}.  We
 fit each experimental correlation function to the form:
\begin{equation}\label{eq:bowler}
C({\mathbf q},\Phi) = N \cdot \bigl[ (1-\lambda)\cdot 1 + \lambda \cdot K({\mathbf q}) \bigl(1 + G({\mathbf q},\Phi) \bigr)\bigr] ,
\end{equation}
where the $(1-\lambda)$  and $\lambda$ terms account for the non-participating and participating 
fractions of pairs, respectively, $K(Q_{\rm inv})$ is the square of the Coulomb wave-function, 
$N$ is the normalization, and  $G({\mathbf q},\Phi)$ is 
the Gaussian form~\cite{pb}:
\begin{equation}\label{eq:prattBertsch}
G({\mathbf q},\Phi) = e^{ - q_o^2 R_o^2(\Phi) - q_s^2 R_s^2(\Phi) - q_l^2 R_l^2(\Phi) - q_o q_s R_{os}^2(\Phi)} .
\end{equation}
$R^2_i$ are the squared HBT radii,
where the $l$,$s$,$o$ subscripts indicate the long (parallel to beam), side (perpendicular to beam and 
total pair momentum) and out (perpendicular to $q_l$ and $q_s$) decomposition of ${\mathbf q}$.

% good until here on citations

\section{Results, Fourier Coefficients and Estimating Eccentricity}

\begin{figure}[t]
%\vspace*{-0.8cm}
\begin{minipage}[b]{65mm}
\centerline{\epsfxsize=60mm\epsfbox{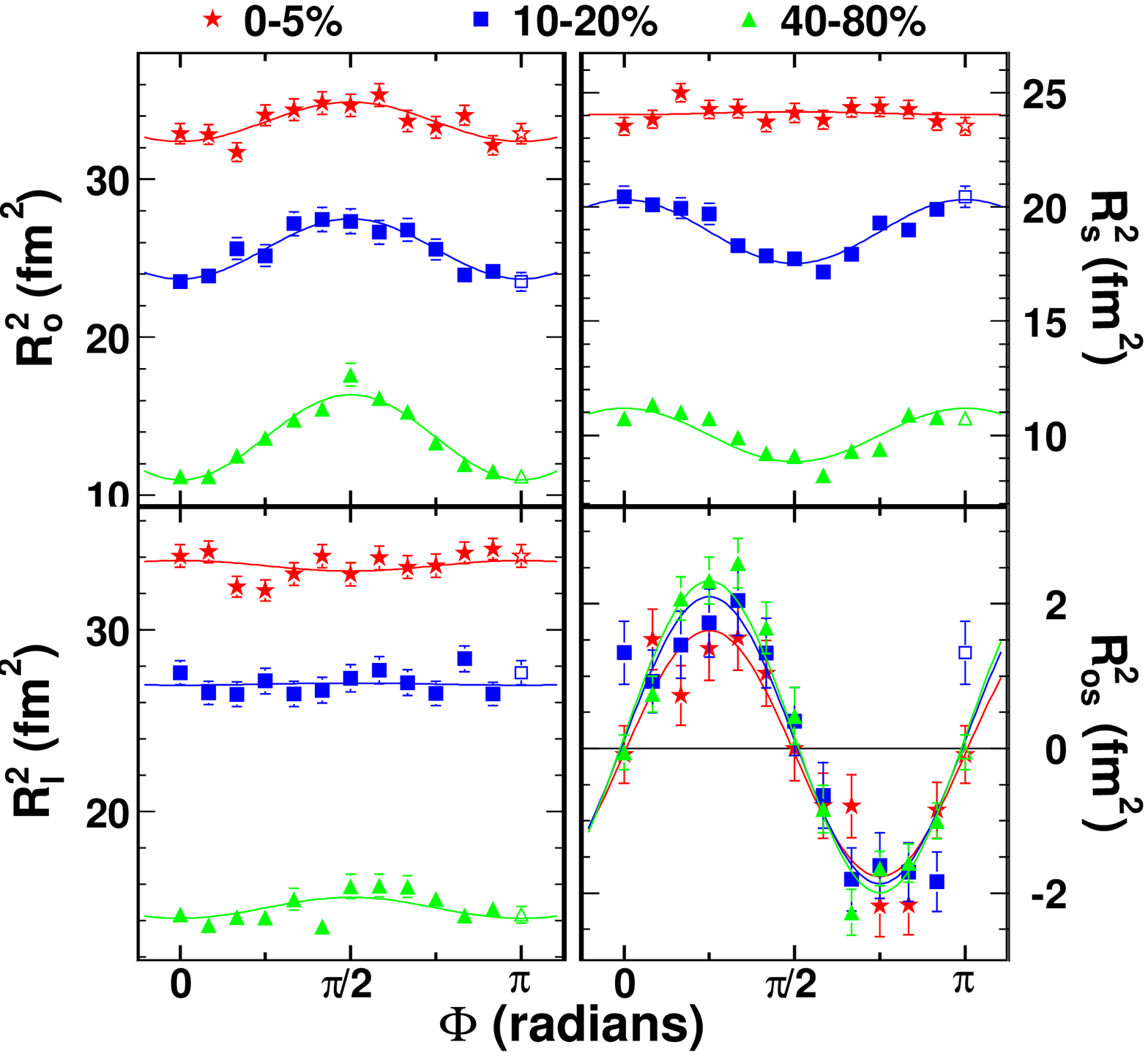}}
\caption{(Above) Squared HBT radii measured
relative
to the reaction plane, for three selections of
event centrality.
%Solid lines show fits to allowed~\cite{HHLW02} oscillations.
\label{fig:STARasHBTfig1}
}
\caption{
\vspace*{0.3mm}
(At right) Blast-wave~\cite{RL03} calculations of Fourier coefficients
for the oscillations of squared HBT radii, for three sources of varying
transverse eccentricity.  Arrows indicate approximation of
Eq.~\ref{eq:eps-approx}.
%   See text for details.
\label{fig:BW3sources}
}
\end{minipage}
\hspace{\fill}
\vspace*{-3cm}
\begin{minipage}[b]{45mm}
\centerline{\epsfxsize=65mm\epsfbox{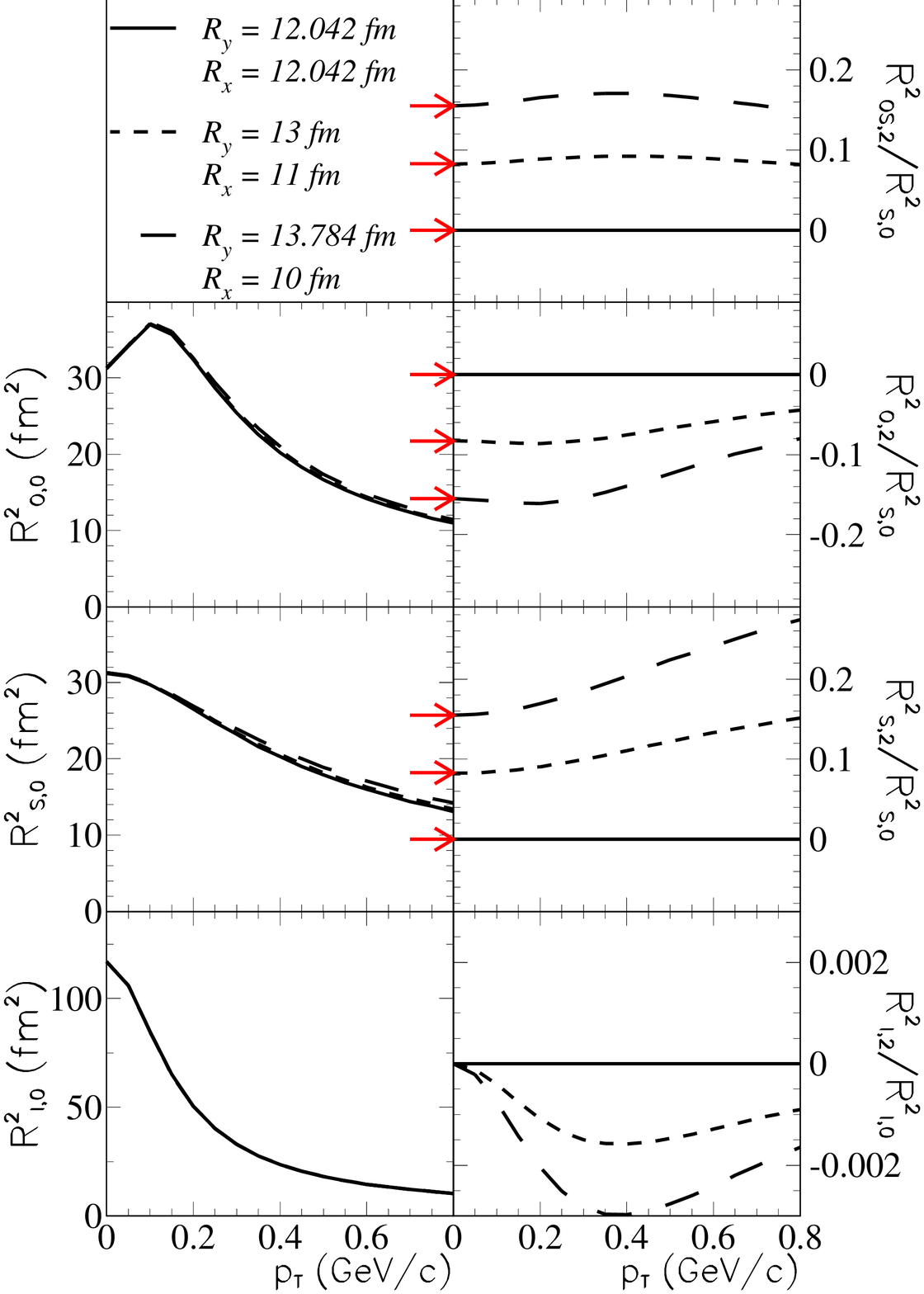}}
\end{minipage}
\vspace*{28mm}
\end{figure}

Figure~\ref{fig:STARasHBTfig1} shows the squared HBT radii as a function of $\Phi$, for three centrality classes and $0.15\leq k_T \leq 0.6$~GeV/c.
Curves correspond to a Fourier decomposition of the squared-radii oscillations according to~\cite{HHLW02,HeinzISMD}
\begin{equation}\label{eq:FC}
R^2_{\mu,n}(k_T) =
\left\{ \begin{array}{ll}
\langle R^2_\mu(k_T,\Phi) \cos(n\Phi) \rangle & (\mu = o, s, l) \\
\langle R^2_\mu(k_T,\Phi) \sin(n\Phi) \rangle & (\mu = os)
\end{array}
\right.
\end{equation}
As expected~\cite{STARy1}, the 0$^{\rm{th}}$-order Fourier 
coefficients (FCs), i.e the mean value of the radii, indicate larger source sizes for more central 
collisions.  We verified that the 0$^{\rm{th}}$-order FC for $R^2_o$, $R^2_s$ and $R^2_l$ correspond to HBT radii obtained in an 
azimuthally-integrated analysis.  Higher-order ($n>2$) FCs were found to be negligible.

%Odd-order (e.g. $n=1$) oscillations, while
%physically interesting~\cite{LHW00}, are inaccessible in this analysis, since we correlate with the second-order event plane~\cite{HHLW02}.

HBT radii directly measure homogeneity regions~\cite{HBTgeneral,Sinyukov_homogeneity}, which constitute only part of the
``whole'' source; hence, connections to the overall source geometry inevitably rely on a model.  The so-called blast-wave
model~\cite{RL03} has been relatively successful in describing soft physics observables (spectra, elliptic flow,
azimuthally-integrated HBT) at RHIC.  In the context of a generalized blast-wave, with spatial and flow anisotropies,
a recent study~\cite{RL03} finds that the transverse source shape is related to $R^2_s(\Phi)$ as
\begin{equation}
\epsilon \equiv \frac{\left( R_y^2 - R_x^2\right)}{\left( R_y^2 + R_x^2\right)} \approx \frac{1}{2}\cdot\frac{R^2_{s,2}}{R^2_{s,0}} 
\approx \frac{1}{2}\cdot\frac{R^2_{o,2}}{R^2_{s,0}} \approx \frac{1}{2}\cdot\frac{R^2_{os,2}}{R^2_{s,0}} .
\label{eq:eps-approx}
\end{equation}
Where $R_x$ and $R_y$ are the source length scales in and out of the reaction plane, respectively.
The approximation becomes exact in the case of vanishing transverse flow~\cite{W98,LHW00}.  In Figure~\ref{fig:BW3sources} are the
$0^{\rm th}$ and $2^{\rm nd}$-order FCs, as defined in Equation~\ref{eq:FC}, for three sources as calculated in
the blast-wave.  The sources have the same temperature, radial flow, timescales, and average size (all adjusted
roughly to the values which best describe RHIC data), but differing $\epsilon$.  
Changing from a round ($\epsilon = 0$) to a deformed source increases the oscillations of the transverse radii (and, very slightly-- note
the scale) the longitudinal radius.  Equation~\ref{eq:eps-approx} arises from the observation that
$2\epsilon$, indicated by small arrows on the right panels, approximately reproduces the relative oscillations in the transverse radii.
Therefore, relying on the blast-wave model for context, we may estimate the freeze-out source shape.

\begin{figure}
\begin{center}
\epsfig{file=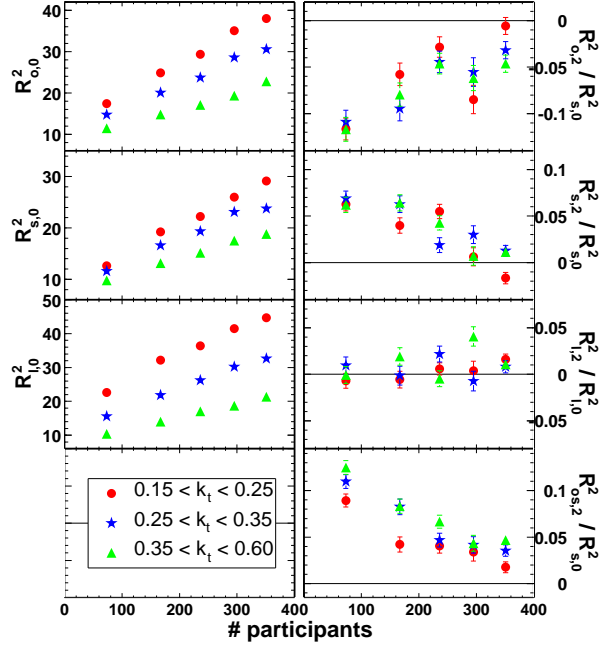,width=8cm}
\end{center}
\caption{Fourier coefficients for the oscillations of squared
HBT radii, for 3 $k_T$ bins, versus centrality
(number of participating nucleons) measured by STAR.
Left panels: means (0$^{\rm th}$-order FC) of oscillations; right panels:
relative 2$^{\rm nd}$-order oscillation amplitudes.  See text for details.
\label{fig:STARasHBTfig3}
}
\end{figure}

Figure~\ref{fig:STARasHBTfig3} shows the measured FCs, as a function of centrality, for three bins in $k_T$, quantifying
the increasing size (left panels) and decreasing anisotropy (right panels) as the collision becomes more central.
In Figure~\ref{fig:STARasHBTfig4}, we compare the initial anisotropy, $\epsilon_{\rm initial}$,
calculated with a Glauber model outlined in reference~\cite{starHighPt}, with $\epsilon_{\rm final}$, estimated by applying
Equation~\ref{eq:eps-approx} to the data of Figure~\ref{fig:STARasHBTfig3}.
We note that $\epsilon_{\rm final}$ has a meaningful sign, and its positive values indicate that the timescales and/or
flow velocities of these collisions are not sufficiently large to fully quench the initial out-of-plane geometric
anisotropy of the system.

\section{A connection to timescales}

\begin{figure}[t]
\vspace*{-0.8cm}
\begin{minipage}[t!]{55mm}
\centerline{\epsfxsize=65mm\epsfbox{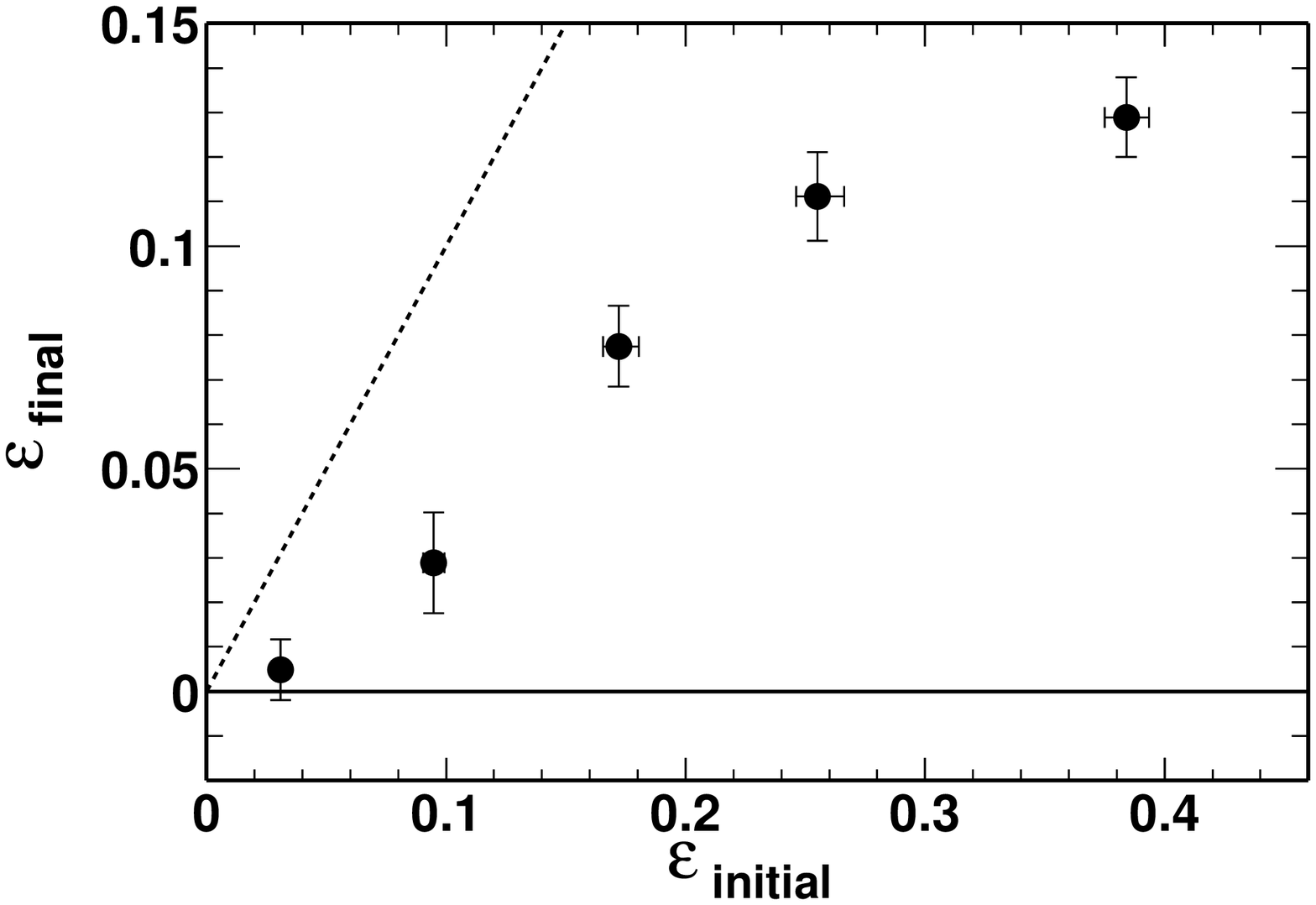}}
\caption{
Final (from HBT) versus initial (from Glauber) transverse source
eccentricity, estimated from the data in Fig.~\ref{fig:STARasHBTfig3}
and Eq.~\ref{eq:eps-approx}.  The line represents
$\epsilon_{\rm initial} = \epsilon_{\rm final}$.
\label{fig:STARasHBTfig4}
}
\end{minipage}
\hspace{\fill}
\vspace*{-3cm}
\begin{minipage}[b!]{55mm}
\centerline{\epsfxsize=60mm\epsfbox{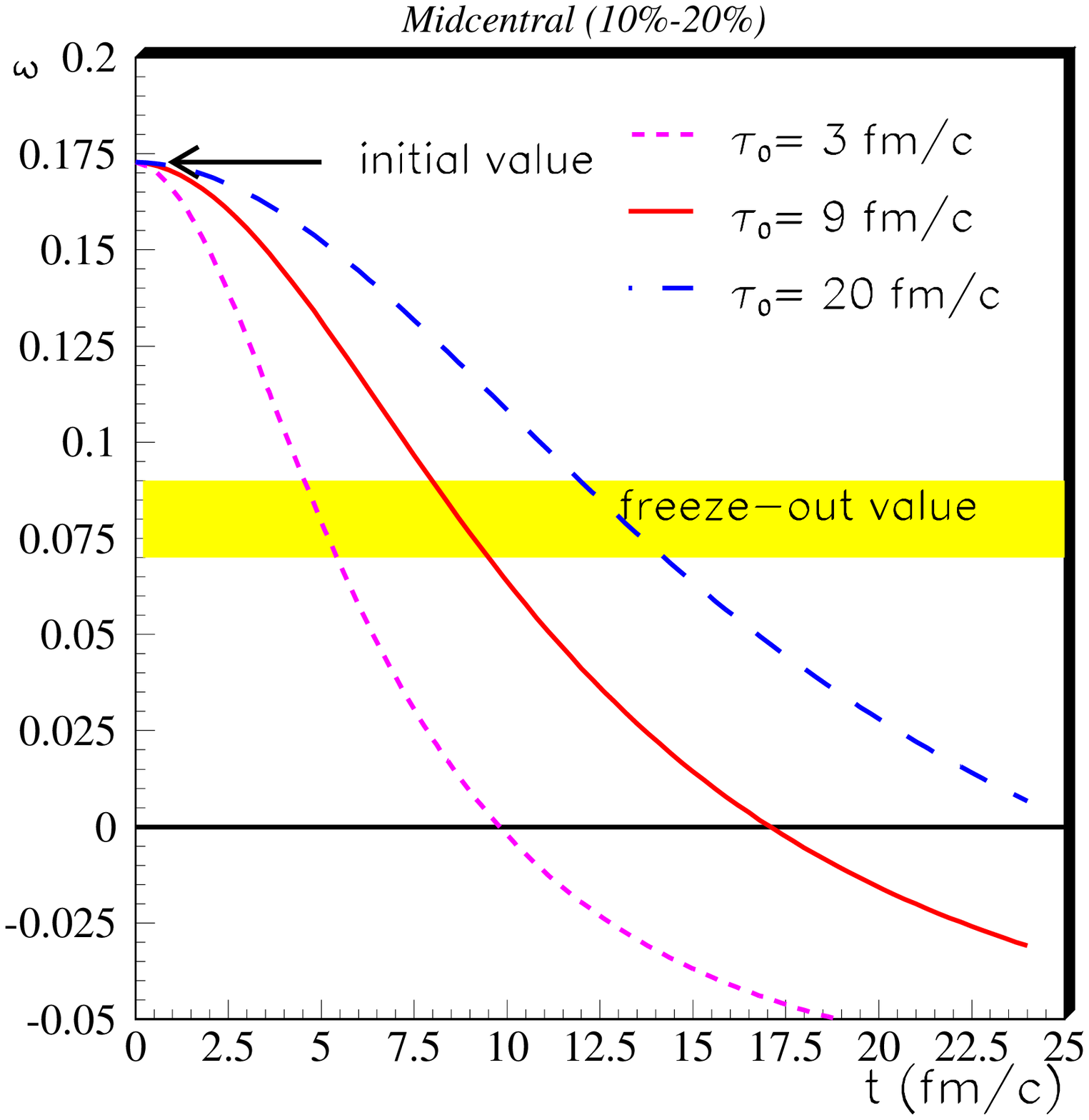}}
\caption{Toy-model calculations for the evolution
of the transverse eccentricity as a function of time.  See text
for details.
\label{fig:EpsEvolution}}
\end{minipage}
\vspace*{26mm}
\end{figure}

The data in Figure~\ref{fig:STARasHBTfig3} should provide useful constraints on timescales and evolution in transport
models of heavy ion collisions.  As a crude example, here we take the model-dependent data in Figure~\ref{fig:STARasHBTfig4}
one step further.
Motivated by 
hydrodynamical calculations~\cite{Kolbv4}, we assume that the transverse flow strength grows roughly linearly with time,
and anisotropies in the flow field set in very early.  Since there is no collective flow initially,
the flow velocity and source extensions in ($x$) and out of ($y$) the reaction plane evolve with time as
\begin{equation}
\beta_{x,y}(t) =  \frac{t}{\tau_0}\cdot\beta_{x,y}^{\rm FO} \quad \longrightarrow \quad
R_{x,y}(t) = R_{x,y}^{\rm Glauber} + \frac{1}{2}\left(\frac{\beta_{x,y}^{\rm FO}}{\tau_0}\right)t^2 .
\label{eq:ToyModel}
\end{equation}
Here, $\beta^{\rm FO}$ is the flow velocity (at the edge of the source) at freeze-out, and $\tau_0$ is the
evolution time of the system-- from initial overlap until freeze-out.  From Equation~\ref{eq:ToyModel}, we may
calculate $\epsilon(t)$ in this toy model.

For collisions between 10-20\% of the cross-section (3$^{\rm rd}$ centrality bin in Figures~\ref{fig:STARasHBTfig3}
and~\ref{fig:STARasHBTfig4}), the estimated initial and final anisotropies are indicated on Figure~\ref{fig:EpsEvolution}.
Estimates for $\beta_{x,y}^{\rm FO}$ were taken from blast-wave fits~\cite{RL03} to mid-central collisions at $\sqrt{s_{NN}}=130$~GeV.
In order to evolve $\epsilon$, we must assume a value for $\tau_0$.  Blast-wave fits~\cite{RL03} to azimuthally-integrated
longitudinal HBT radii suggest a rather short evolution time of $\sim 9$~fm/c.  Assuming this value, we arrive at a consistent
picture: as shown by the solid line in Figure~\ref{fig:EpsEvolution}, at $t=\tau_0=9$~fm/c, $\epsilon$ has reached its
measured value.  On the other hand, using a significantly different value for $\tau_0$ yields inconsistent results, as
shown by the dashed and dotted lines.  Thus, our measurements, in the context of this admittedly crude model, seem to
provide corroborating evidence in the direction of evolution timescales on the order of 9~fm/c.

\section{Summary}

We have presented an analysis of azimuthally-sensitive pion interferometry for Au+Au collisions at RHIC.
We observe clear 2$^{\rm nd}$-order oscillations of the transverse HBT radii, and extract Fourier coefficients.
Geometrically, as the impact parameter decreases, the $0^{\rm th}$-order FCs increase, indicating larger sources,
and the relative oscillations decrease, suggesting rounder ones.  More quantitative statements rely on model
interpretations.  

In the blast-wave model, the relative oscillations provide a good estimate of the source
ellipticity at freeze-out.  In this context, we find that the source at all centralities is extended out of
the reaction plane (though not as much as the initial source), suggesting rather short evolution durations.

This data should provide strong constraints on dynamic models of heavy ion collisions, particularly as
regards timescales and pressure anisotropies.  To illustrate the point, we performed a toy model calculation
which turned out to favor the rather short evolution timescales extracted via other means previously.

\end{document}